\documentclass[fontsize=11pt, onecolumn]{article}	 

\usepackage[english]{babel} 
\usepackage[protrusion=true,expansion=true]{microtype} 
\usepackage{amsmath,amsfonts,amsthm} 
\usepackage[svgnames]{xcolor} 
\usepackage[hang, small,labelfont=bf,up,textfont=it,up]{caption} 
\usepackage{booktabs} 
\usepackage{fix-cm}	 

\usepackage{sectsty} 
\allsectionsfont{\usefont{OT1}{phv}{b}{n}} 

\usepackage{fancyhdr} 
\usepackage{lastpage} 

\usepackage[square, comma, sort&compress, numbers]{natbib}
\usepackage{url}
\usepackage{breakurl}
\usepackage[breaklinks=true,colorlinks=true,linkcolor=blue,urlcolor=DarkBlue,citecolor=DarkBlue]{hyperref}
\usepackage{graphicx}

\usepackage{lettrine} 
\newcommand{\initial}[1]{ 
\lettrine[lines=4,lhang=0.3,nindent=0em]{
\color{DarkGoldenrod}
{\textsf{#1}}}{}}

\newcommand{\eg}{\hbox{\sl e.g.},}

\usepackage{titling} 

\newcommand{\HorRule}{\color{DarkGoldenrod} \rule{\linewidth}{1pt}} 

\pretitle{\vspace{-50pt} \begin{flushleft} \HorRule \vspace{16pt} \fontsize{24}{24} \usefont{OT1}{phv}{b}{n} \color{DarkBlue} \selectfont} 

\title{Persistent, Global Identity for Scientists via ORCID }

\posttitle{\par\end{flushleft}\vskip 0.5em} 

\preauthor{\begin{flushleft}\large \lineskip 0.5em \usefont{OT1}{phv}{b}{n} \color{DarkBlue}} 

\author{August E. Evrard$^{1, 5}$, Christopher Erdmann$^{2, 5}$, Jane Holmquist$^{3,5}$, James Damon$^{2}$, Dianne Dietrich$^{4}$}

\postauthor{ \newline \newline  \footnotesize \usefont{OT1}{phv}{m}{sl} \color{Black} 
$^1$University of Michigan, Ann Arbor, MI \& Institut d'Astrophysique, Paris, France\\
$^2$Harvard-Smithsonian Center for Astrophysics, Cambridge, MA\\
$^3$Princeton University Library, Princeton, NJ\\
$^4$Cornell University Library, Ithaca, NY\\
$^5$ORCID Ambassador

\par\end{flushleft}\HorRule} 

\date{} 

\begin{document}

\maketitle 

\thispagestyle{fancy} 


\initial{S}\textbf{cientists have an inherent interest in claiming their contributions to the scholarly record, but the fragmented state of identity management across the landscape of astronomy, physics, and other fields makes highlighting the contributions of any single individual a formidable and often frustratingly complex task.  The problem is exacerbated by the expanding variety of academic research products and the growing footprints of large collaborations and interdisciplinary teams.  In this essay, we outline the benefits of a unique scholarly identifier with persistent value on a global scale and we review astronomy and physics engagement with the Open Researcher and Contributor iD (ORCID) service as a solution. }


\section{Introduction} 

The scientific record reflects an enduring conversation among researchers around the globe and across the millenniums,   
from Aristotle and Ptolemy through Maxwell and Einstein and up to the present. 
Each day, new voices join the conversation and new ideas are expressed in articles, books, software, data and other products created by scientists and conserved by libraries and other repositories \citep{Hilton2013}.  

Digital technologies have accelerated the pace of this discussion.\footnote{Consider the 2014 claimed discovery of primordial gravity waves in cosmology, where theoretical papers \citep[\eg][]{Nakayama1403.4132, Harigaya1403.4536} interpreting the BICEP2 observations appeared on arXiv.org within hours of the experimental results themselves \citep{BICEP2paperI}. }
Open access repositories such as arXiv.org 
provide liquidity for the marketplace of ideas, and public social media outlets  
support open, informal exchanges among colleagues in real time.  Scientists find value in the efficiency of such digital-only discourse, even if preservationists may fret over its fragility \citep{Lynch2003institutional}.  

The conduct of scholarship relies on the ability to register, identify, and persistently access particular snippets of this massive conversation \citep{Branin1998}.  Digital object identifiers \citep[DOIs,][]{DOI, DataCite} help identify \textit{what} was said by providing unique, persistent web links to refereed papers, source materials, data, codes, videos, and many other types of content.  

Identifying \textit{who} contributed to the conversation is much harder \citep{Enserink2009, Warner2010, Fenner2011, Kurakawa2014}.  While journal papers and other media list contributing author names, typically with institutional affiliations, this information is often insufficient to uniquely identify any single individual, especially over time.  The community of scientific researchers -- indeed, the scholarly community as a whole -- lacks  the functional equivalent of a DOI for scholarly identity, a persistent identifier that uniquely links any researcher to her or his body of contributions.  

This is not to say that scholarly identity systems don't exist today.  Anyone who has published in a major journal, posted to arXiv.org, applied for science agency funding, or added content to an institutional or public repository has an identifier for that particular platform. The problem is that such balkanized identities make it nearly impossible to form a complete view of a person's contributions.  In addition, for-profit companies that host scholars' identities may seek to monetize that content in ways that run counter to the open culture of scientific inquiry.  

In this article, we review the benefits to researchers, particularly those early in their careers, of possessing a unique, non-commercial identifier that breaks through existing silos.  After reviewing the fragmented landscape of scholarly identity in \S\ref{sec:landscape}, we describe the Open Researcher and Contributor iD (ORCID) in \S\ref{sec:ORCID} and discuss ongoing integration efforts relevant to the domains of astronomy and physics.  We briefly discuss ORCID in a broader context of Internet identity in \S\ref{sec:discussion} before summarizing in \S\ref{sec:summary}. 

We do not attempt to offer general advice on managing a professional profile \citep[though for some concise guides, see][]{Markgren2011, ImpactStoryGuide2014}.  Nor do we venture into credentialing, the realm of authorization and authentication to Internet services \citep[\eg InCommon,][]{InCommon}, although potential ``cross-walking'' applications are mentioned below.  
Rather, we seek to motivate the case for linking each researcher with the scholarly products created by her or him in a unique, persistent, and global manner, and present ORCID as a solution.   

The authors of this paper include three ORCID ambassadors, a group who seeks to ``encourage the adoption of ORCID identifiers locally and globally'' \cite{ORCIDambassadors}.   




\section{A Complex Identity Landscape}\label{sec:landscape}

Each of us is accustomed to having a unique civilian identity based on a valid birth certificate, passport, or social security number issued by a governmental authority.  In contrast, beyond degree certification documents, professional scholars have no uniform mechanism to assign a unique identifier to an individual.   Instead, layers of tangled identities develop around us.  

\subsection{Layers of identities}

\begin{figure}[t]
\begin{center}
\vspace{-1.0truecm}
\includegraphics[width=1.05\columnwidth]{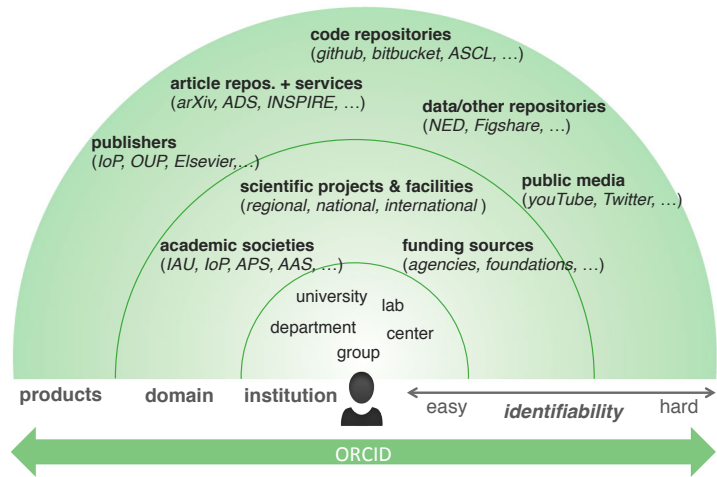}
\vspace{-3.0truecm}
\caption{A schematic view of layers across which a researcher typically operates using multiple identities.  The innermost layer involves identity at one's host {\bf institution} while a {\bf domain} layer involves interactions with professional organizations, projects and societies.  The {\bf products} layer adds publishing houses and hosting services for articles, data, code, and other scholarly products.  The broadening scope and scale makes it harder to uniquely identify an individual in the outer layers.   Uniting the layers with a single scholarly identifier -- an ORCID iD --  clarifies the connections between repository items and the projects, funding sources, and individuals responsible for creating them.  
\label{fig:landscape}
}
\end{center}
\end{figure}

Researchers in astronomy, physics or similar disciplines work in a larger ecosystem of scholarly endeavor that we present as a set of three layers in Figure~\ref{fig:landscape}.  This representation is simplistic, and not intended to be complete, but it serves to separate the organizations that scientists typically interact with into different functional roles.  

The \textit{institution} layer consists of a researcher's home institution and its internal components.  At this level, digital identity is, by and large, managed by the central information technology organization of the organization.  Even so, multiple identities can exist here due to internal systems that have not been centrally aligned or due to free-floating services (wikis, databases, etc.) that host their own authentication.  
 
Outside the institution, a \textit{domain} layer includes organizations that serve and fund science, as well as the facilities and collaborations that generate new discoveries.  Interacting with the information systems in this layer typically requires creating separate identities for each organization, although an institutional e-mail address often serves as a common naming device.  Funding agencies and most scientific societies are active at the national level, whereas large projects and facilities, and transnational  societies such as the International Astronomical Union, support members from around the globe.  The international nature of science demands a global identity solution.   

An outer \textit{products} layer holds the materials generated within the inner two layers.  This space contains publishing in both traditional and modern forms, the former emphasizing commercial publishing concerns and the latter including a variety of web publishing formats, many of them open access.  Traditional companies such as Reed-Elsevier and Thomson-Reuters have developed their own mechanisms for identifying individual authors, and these identities inform their commercial research analytics platforms Scopus \citep{Scopus} and Web of Science \citep{WebOfScience}, respectively.  

While peer-reviewed literature published in established journals remains a dominant force across physics and astronomy, the open market created by arXiv.org rivals in importance because it provides real-time access to new science.  The arXiv, which displaced the distribution of physically mailed pre-prints with an open-access e-print repository, is complemented by the abstract and indexing services provided by INSPIRE \citep{INSPIRE} and ADS \citep{ADS}.  These sites emerged in the early days of the World-Wide Web and are now heavily used by astronomers and physicists as communication, discovery and assessment tools.   


Newer services in the products layer host and distribute code, data, and other digital products such as presentation slides, videos, and lab notebooks.  Some sites focus on the needs of a particular science domain while others, including the new non-commercial service Zenodo \citep{Zenodo} or commercial software repositories such as \textit{Github} \citep{github} or \textit{Bitbucket} \citep{bitbucket}, serve a much broader audience beyond astronomers and physicists.  

The landscape charted in Figure~\ref{fig:landscape} is by no means exhaustive, but most practicing astronomers and physicists at any rank should find its elements familiar.  Accrediting and linking contributions  across the entire landscape of Figure~\ref{fig:landscape} is a rational desire that would benefit both individuals and organizations.  From that basic concept, we now turn to offer a few practical reasons to care about a unique scholarly identity.  

\subsection{Practical motivations for unique scholarly identity }

Scientists are professionals who need to advance their careers, help deserving colleagues move forward in their careers, and generally take part in the larger workings of scientific society that includes evaluating, crediting and criticizing the work of others.  Here are some common activities that a unique identifier can help support.   

\begin{itemize}

\item[o]\textbf{Claim appropriate credit.}  For two people in the same field sharing a name, the opportunity for disambiguation serves as the simplest motivation for a unique identifier.  The pair of Jens Jensens at the Niels Bohr Institute in Copenhagen would certainly benefit.  So too would the more than 100 Y. Zhangs who are registered to arXiv.org \citep{Warner2010}.   The difficulty in identifying and evaluating publications authored by someone with a common name often frustrates writers of recommendation letters.  

\item[o]\textbf{Manage life transitions.}   Circumstances such as marriage can lead to a change in one's name.  In the present, fragmented state of affairs, the responsibility is entirely on the researcher to inform all the elements in the academic landscape of such a change.  How easy will it be for others to view the full body of work for someone experiencing this personal transition?  Will discovery services like ADS and INSPIRE automatically extend searches across the new and old names?   A unique ID associated with both names is a simple mechanism to achieve continuity through such transitions.  

\item[o]\textbf{Manage career migration.}  Researchers in the early stages of an academic career are likely to change institutional affiliation multiple times as their careers advance (as nicely illustrated for dotAstronomy members by Daniel Foreman-Mackey \citep{dotAstronomy2014}).  Some will exit the academic profession and transition to other employment.  A persistent identifier that follows a person through these career transitions would allow that individual's contributions to the products layer to be easily consolidated over time and across multiple locations, including those in industry or commerce.  

\item[o]\textbf{Be socially responsible.}  Linking scientific output to taxpayer investment is a socially responsible thing to do.  In 2013, the Office of Science and Technology Policy in the United States directed federally-funded agencies ``to support increased public access to the results of research funded by the Federal Government'' \citep{OSTPmemo2013}, and this mirrors a similar announcement by the European Commission \citep{ECmemo2012}.  A collection of repositories of open access products tagged with ORCID identifiers constitutes a knowledge fabric capable of supporting these directives by facilitating links between materials in the products layer, funding sources and projects in the domain layer, and grant recipients within institutions.  

\item[o]\textbf{Stand out in a crowd.}  Scientific research is increasingly collaborative \citep[\eg][]{Adams2005}, but working in large teams on long-term projects is viewed as potentially detrimental to early career scientists.  Will a graduate student's contributions to an alphabetically-ordered publication with a hundred or more authors be appropriately recognized?   While the promotion and tenure process still relies heavily on letters from expert reviewers \citep{Brand2013}, so-called ``arms-length''  letter writers may themselves be unaware of an individual's specific contributions.  For projects that operate transparently in the products layer, a unique author ID that labels the codes built, data sets created, and presentations made by a young researcher can bring those many contributions into focus.  


\end{itemize}


\section{Open Researcher and Contributor iD (ORCID)}\label{sec:ORCID}

In 2010, a group of publishers, academics, and librarians incorporated a non-profit organization with a mission to solve the name ambiguity problem in scholarly communication.   The Open Researcher and Contributor iD \citep{ORCID} solution takes the form of a central registry of unique identifiers, known as ORCID iDs, for individual researchers along with open and transparent mechanisms to link these with existing researcher identifier schemes.  ORCID principles \cite{ORCIDprinciples} 
 emphasize openness, transparency, broad scholarly scope on a global scale, and respect for the right of researchers to set privacy settings on their records.   

The service is governed by a board of sixteen members with representatives from multiple areas including publishing, academia and information science.  The American Physical Society was a launch partner of ORCID, and the American Astronomical Society joined as a sponsor in 2011.  Similar to arXiv, a member fee model supports the organization's operations.  

\subsection{ORCID Services}

The basic service that ORCID provides is the creation of a unique, 16-digit identifier for an individual.  The cost of entry is low; a name and e-mail address is all that's needed to create an identifier, with additional information such as affiliation and position optional.  After an ORCID iD is created, the owner has choices to make regarding populating and sharing additional information in the ORCID registry.  Ideally, all work across the landscape of Fig.~\ref{fig:landscape} would be rounded up and used to populate one's registry, but this goal is only partly possible today.  We discuss below the status of some automatic ingest capabilities for physics and astronomy.   

There are three privacy settings --- public, limited access, and private --- that can be applied in a granular manner to registry contents.  Settings are chosen at the time of account creation and can be modified at any time.  A  researcher can choose to become ``invisible'' to the public by marking all fields other than the ORCID identifier as private.  ORCID does not require a researcher to populate and expose registry contents through the orcid.org website.  

Account delegates may be assigned to manage a registry.  The application program interfaces for ORCID services allow account management to be automated, and one can control the set of trusted parties that are able to update registry contents.  

A large, active list of integration efforts is maintained on the web site's integration page \citep{ORCIDintegration}.  Efforts relevant to the astronomy and physics communities 
are summarized in Table~\ref{tab:engagement}.   

\subsection{ORCID integration at academic institutions}

Dozens of universities across the world have enrolled as ORCID members.  In the US,  both private (\eg\ Boston University, Caltech, Harvard, and MIT) and public (\eg\ Texas A\&M University, Purdue University, and University of Michigan) institutions are at various stages of planning and implementing the assignment of ORCID iDs to their researchers.  

The John G. Wolbach Library is actively promoting ORCID within the Harvard-Smithsonian Center for Astrophysics and the larger astrophysics community.   The Texas A\&M University Library began enrolling junior researchers, mainly graduate students and postdocs, with the aim of tagging all doctoral dissertations with ORCID iDs.  The University of Michigan Library is assigning ORCID iDs within the School of Medicine in 2014, with other colleges to follow.  
 

\begin{table}
\begin{center}
\begin{tabular}{ll}
\hline
\textbf{Organization}  & \textbf{Activity} \\   
\hline
Various universities  & Assigning ORCID iDs to campus researchers   \\            
NSF, DoE, NASA  & Planning for ORCID iDs in grant management \\            
AAS, APS  & Using ORCID iDs for journal authors, referees \\            
INSPIRE, ADS & Developing ORCID iD integration services \\
Zenodo, {\sl Github}, Figshare & Support ORCID iD tags for data and code  \\

\hline
\end{tabular}
\caption{ORCID engagement within the physics and astronomy communities.}\label{tab:engagement}
\end{center}
\end{table}

\subsection{ORCID in the domains of physics and astronomy}

Within the US, the American Physical Society (APS) and American Astronomical Society (AAS) have endorsed ORCID as a mechanism for solving author disambiguation and for streamlining the workflow of scholarship.   APS members can apply for an ORCID via the APS Author Profile page \citep{APSprofile}.  The AAS has invited its members to add ORCID iDs to their profile \citep{AASprofile} as part of a larger digitization strategy that is planned to include DataCite DOIs for data associated with  AAS-sponsored journal articles \citep{AASdigital}.  The American Institute of Physics and UK-based Institute of Physics are also ORCID members. 

Science agencies across multiple countries, including Japan, the Netherlands, Portugal, the UK and the US have 
called for the use of persistent identifiers for researchers involved in grants submissions and management processes.  In the US, adoption of ORCID iDs is being led by the National Institutes of Health \citep{NIH-ORCID2014}, and other agencies are at various stages of planning.  
The Office of Scientific and Technical Information in the US Department of Energy (DoE), the part of the agency that maintains and disseminates the results of DoE's \$10 billion annual investment in research and development, joined ORCID as a member in May 2013.  The US National Science Foundation sees potential benefits in unique identifiers, including strengthening its portfolio analysis capabilities, facilitating the process of research reviews, and reducing administrative burdens \citep{ORCID-NSF-OCT13}.  

\subsection{ORCID in the products layer}

The variety of elements that comprise the products layer continues to expand as the digital era of scholarship matures.  Indeed, the notion of \emph{scholarly products} as a term to encompass contributions beyond traditional publication forms of journal articles, conference proceedings, and books is itself relatively new.  Platforms that host the varied outputs of science are now a mix of commercial and open access services, the latter including repositories hosted by research universities.  Due to its complexity and evolving nature, it is difficult to comprehensively review ORCID engagement across this layer;  we provide here a snapshot of activities relevant to the needs of astronomers and physicists.


In terms of article repositories, ORCID currently supports imports from the commercial indexing services SCOPUS and Web of Science.  These services focus on tracking products in peer-reviewed journal literature.  On the other hand, physicists and astronomers mainly use non-commercial services for search and discovery, namely INSPIRE and ADS, respectively.  ORCID integration with these services is under development.  


INSPIRE, managed by CERN, currently allows author profiles to be linked with ORCID, and it will soon deploy tools to align the ORCID registry with INSPIRE records via push/pull operations.  INSPIRE will retrieve any missing publications from ORCID, and an author claiming system will allow the addition of INSPIRE records to an ORCID registry.   As of June 2014, roughly 250 INSPIRE authors have linked ORCID iDs to their records (Laura Rueda, priv. comm.).    APS reports over 5500 authors and more than 3200 referees registered with ORCID iDs (Mark Doyle, priv. comm.).  The latter constitutes roughly 10 percent of the overall referee cohort. 

For astronomers, the NASA ADS service is in the planning stage of ORCID integration (Alberto Accomazzi, priv. comm.).  As of June 2014, over 1600 authors on ADS have ORCID iDs (Carolyn Grant, priv. comm.), and 11\% of authors in AAS journals have ORCID iDs (C. Biemesderfer, priv. comm.).  


The arXiv.org e-Print archive is also moving towards ORCID integration. The 2014 arXiv Roadmap \citep{arXivRoadMap} lists support for ORCID identifiers as a goal and notes improved interoperability with authority records from external repositories and better mechanisms for reporting institutional statistics to member organizations as benefits. 



Beyond articles, ORCID iDs can be integrated into {\sl Github} as part of a new collaboration with the Zenodo 
service.  Zenodo serves as ``an open digital repository for everyone and everything that isn't served by a dedicated service; the so called `long tail' of research results'' \citep{Zenodo}.  The ODIN project \citep{ODIN} is building a tool that will allow ORCID iDs to be added to research products associated with DataCite DOIs.  The HubZero science collaboration platform \citep{HubZero} is also integrating ORCID iDs into its workflows.  
ORCID identifiers for content creators can be included in the citation metadata for datasets in the DataCite Metadata Store.  The public hosting service Figshare \citep{Figshare} allows users to tag their entries with ORCID iDs.   

In short, a modest but growing level of ORCID integration exists in the sector of the products layer relevant to astronomers and physicists.  




\section{Discussion}\label{sec:discussion}


We have emphasized the benefits of ORCID iDs but there are associated costs and potential risks.  In a 2014 discussion about ORCID on the Astronomy Facebook page, some participants  expressed skepticism about having yet another on-line identity to maintain, especially if its use is mandated by some organization.   While there is a time cost for an individual to obtain an iD, it can be reduced to under five minutes by setting one's privacy settings so that all registry content is private.  This approach avoids having the ORCID web site become yet another professional site one needs to maintain.  

The dominant sentiment of the Astronomy Facebook discussion, however, was positive, with the most liked response coming from a Japanese researcher whose name is easily confused with others.  He wrote, 
\begin{quote} 
\textsl{It just doesn't make sense to have to wonder about completeness and ambiguity when looking up someone's scientific record.}
\end{quote} 
Frustration of this kind will be eliminated by wide ORCID adoption. 

As a young organization attempting to solve a complex problem on a global scale, ORCID has its share of growing pains.  Some recent negative scrutiny surrounded an apparent inability to opt out of Google Analytics \citep{GoogleAnalytics} tracking on orcid.org, but this issue was found to a technical fault with the specific blocking software being used \citep{ORCIDissue}.  

In fact, ORCID membership is rapidly growing.  In mid-November 2014,  two years 
after the launch of the ORCID Registry, the organization issued its one millionth identifier.  Versions of the ORCID site in Russian and Portuguese soon followed.  

\subsection{ORCID in the broader context of Internet identity }

It is important to recognize that ORCID augments rather than replaces the balkanized set of identifies associated with the landscape of Figure~\ref{fig:landscape}.  Individual institutions such as universities and national laboratories will continue to maintain identity and access management services based on local identities.  But, as Keith Hazelton of Internet2 and University of Wisconsin points out, ORCID iDs could be used as an identifier to ``cross-walk'' among appropriately federated campus systems \citep{Hazelton2014}.   Such a process could make collaborative research more efficient by simplifying access to pools of data and computing resources that cross institutional boundaries.   

As argued in the Astronomy Facebook page discussion, the registry aspect of ORCID is a mixed blessing.  On one hand, a site that automatically populates itself with the papers published, grants received, codes authored, etc., by an individual sounds like a great idea.  On the other hand, INSPIRE and ADS already provide such a service for publications, and most researchers already expose their professional profiles through personal web pages and/or existing commercial services offered by the likes of linkedin.com, researchgate.com, mendeley.com, and academia.edu.  How does the orcid.org registry relate to these other profile solutions?   

This is a simple question without a simple answer, but we stress that ORCID principles emphasize personal control of one's registry content, so there are a number of possible paths.  As already mentioned, there is the simple option of declaring an ORCID profile private which minimizes its exposure on orcid.org.  As INSPIRE and ADS integration with ORCID advances, the option of exposing publications through orcid.org should become more attractive.  The ability to claim a broader variety of products such as data, codes, videos and blog posts across a set of repositories is also an appealing aspect.  

Finally, we note that there is an independent effort supported by the International Standards Organization (ISO) that also seeks to create global identifiers for creative activities.  The International Standard Name Identifier \citep[ISNI,][]{ISNI} has a mission to ``assign to the public name(s) of a researcher, inventor, writer, artist, performer, publisher, etc. a persistent unique identifying number in order to resolve the problem of name ambiguity in search and discovery.''   ISNI and ORCID use the same 16-digit identifier format. In early 2014, the two organizations signed a memorandum of understanding (MOU) to develop a strategic partnership \citep{ISNI-ORCID}.  The following paragraphs from the MOU help differentiate the two organizations.
\begin{quote}
The ISNI identifier is a global standard for identifying the millions of contributors to creative works and those active in their distribution, including writers, artists, creators, performers, researchers, producers, publishers, aggregators, and more across the global information supply chain. It is part of a family of international standard identifiers that includes identifiers of works, recordings, products and right holders in all repertoires, \eg\ DOIs, ISAN, ISBN, ISRC, ISSN, ISTC, and ISWC.
\smallskip

ORCID is an open and interdisciplinary community-based effort to provide a self-claim registry of unique researcher identifiers. To ensure the identifier links researchers with their works, ORCID works with the research community to embed these identifiers in workflows, such as manuscript submission, grant application, and dataset deposition. ORCID is unique because of its direct relationship with researchers and with organizations throughout the research community.
\end{quote}
Simply (if somewhat imprecisely) put, ISNI is aimed at people pursuing broad creative activities while ORCID is more  focused toward academic researchers.  ORCID's existing partnerships with organizations in physics, astronomy, biology and other science and engineering areas makes it the natural choice for researchers in these domains.  The MOU states that ORCID and ISNI ``are agreed on the need to aim for interoperation through linking identifiers and sharing public data between the two systems,''  so it is likely that an ORCID iD will effectively serve as an ISNI in the near future.  

\section{Summary}\label{sec:summary}

Science is increasingly multidisciplinary and collaborative at an international scale.  As a result, discovery, attribution and recognition of individual accomplishments is getting harder.  

A solution to unique, persistent scholarly identity is required, and the ORCID service is now available as an open, non-commercial, cross-disciplinary, global solution.  The learned societies of physics and astronomy have endorsed ORCID from its formative stages and are actively engaged with its adoption.  The benefits to scientific society of such a service are too valuable to ignore, and a solution is ready.

\bigskip
\bigskip

\noindent
{\sl Acknowledgements.}  We thank Kayleigh Ayn Boh\'emier for useful suggestions on an earlier draft of this paper.  AEE acknowledges support from the Marie de Paris Research in Paris program and the Institut d'Astrophysique for a sabbatical visit during which first drafts of this paper were written. 

\bigskip

\ifx\undefined\bysame
\newcommand{\bysame}{\leavevmode\hbox to\leftmargin{\hrulefill\,\,}}
\fi


\end{document}